\documentclass[english]{elsarticle}
\usepackage[T1]{fontenc}
\usepackage[latin9]{inputenc}
\usepackage{amsmath}
\usepackage{graphicx}

\makeatletter
\@ifundefined{date}{}{\date{}}
\@ifundefined{definecolor}
 {\usepackage{color}}{}
 \usepackage{ae}
\makeatother

\makeatother

\usepackage{babel}
\begin{document}

\title{ZE3RA: The ZEPLIN-III Reduction and Analysis Package}

\author[LIP,ICL]{F.~Neves\corref{cor1}}

\cortext[cor1]{Corresponding author: }

\ead{neves@coimbra.lip.pt}

\author[ITP]{D.Yu.~Akimov}

\author[ICL]{H.M.~Araújo}

\author[EDI]{E.J.~Barnes}

\author[ITP]{V.A.~Belov}

\author[ITP]{A.A.~Burenkov}

\author[LIP]{V.~Chepel}

\author[ICL]{A.~Currie}

\author[LIP]{L.~DeViveiros}

\author[RAL]{B.~Edwards}

\author[EDI]{C.~Ghag}

\author[EDI]{A.~Hollingsworth}

\author[ICL]{M.~Horn}

\author[RAL]{G.E.~Kalmus}

\author[ITP]{A.S.~Kobyakin}

\author[ITP]{A.G.~Kovalenko}

\author[ICL]{V.N.~Lebedenko}

\author[LIP,RAL]{A.~Lindote}

\author[LIP]{M.I.~Lopes}

\author[RAL]{R.~Lüscher}

\author[RAL]{P.~Majewski}

\author[EDI]{A.St\,J.~Murphy}

\author[RAL]{S.M.~Paling}

\author[LIP]{J.~Pinto da Cunha}

\author[RAL]{R.~Preece}

\author[ICL]{J.J.~Quenby}

\author[EDI]{L.~Reichhart}

\author[LIP]{S.~Rodrigues}

\author[EDI]{P.R.~Scovell}

\author[LIP]{C.~Silva}

\author[LIP]{V.N.~Solovov}

\author[RAL]{N.J.T.~Smith}

\author[RAL]{P.F.~Smith}

\author[ITP]{V.N.~Stekhanov}

\author[ICL]{T.J.~Sumner}

\author[ICL]{C.~Thorne}

\author[ICL]{R.J.~Walker}

\address[LIP]{LIP--Coimbra \& Department of Physics of the University of Coimbra,
Portugal}

\address[ICL]{High Energy Physics group, Blackett Laboratory, Imperial College
London, UK}

\address[ITP]{Institute for Theoretical and Experimental Physics, Moscow, Russia}

\address[EDI]{School of Physics \& Astronomy, University of Edinburgh, UK}

\address[RAL]{Particle Physics Department, STFC Rutherford Appleton Laboratory,
Chilton, UK}
\begin{abstract}
ZE3RA is the software package responsible for processing the raw data
from the ZEPLIN-III dark matter experiment and its reduction into
a set of parameters used in all subsequent analyses. The detector
is a liquid xenon time projection chamber with scintillation and electroluminescence
signals read out by an array of 31 photomultipliers. The dual range
62-channel data stream is optimised for the detection of scintillation
pulses down to a single photoelectron and of ionisation signals as
small as those produced by single electrons. We discuss in particular
several strategies related to data filtering, pulse finding and pulse
clustering which are tuned to recover the best electron/nuclear recoil
discrimination near the detection threshold, where most dark matter
elastic scattering signatures are expected. The software was designed
assuming only minimal knowledge of the physics underlying the detection
principle, allowing an unbiased analysis of the experimental results
and easy extension to other detectors with similar requirements. \end{abstract}
\begin{keyword}
ZEPLIN-III, liquid xenon detectors, dark matter, signal analyses,
data reduction
\end{keyword}
\maketitle

\section{Introduction\label{sec:Introduction}}

The present work evolved within the ZEPLIN-III experiment, a two-phase
(liquid/gas) xenon detector aiming to measure very low energy nuclear
recoils produced by the interaction of dark matter WIMPs (Weakly Interactive
Massive Particles) \citep{sumner01,araujo:2006aa,Akimov:2007aa,Lebedenko:2008aa}.
Searches for rare particle interactions in low-background physics
experiments are inherently difficult, and any data reduction and analysis
software must address two main challenges. Firstly, unusual event
topologies will almost inevitably appear in the long exposures required
for the science acquisitions (lasting typically for many months or
even years). These are not exercised by the various calibration runs
and may result from localized instabilities or unexpected backgrounds,
for example. Secondly, WIMP-nucleus interactions are expected to result
in very small energy transfers to xenon atoms, and so the search for
scintillation and ionization signatures from WIMPs goes down to the
quantum of response in these channels, i.e. to the photoelectron level
in scintillation and to single electrons in ionization. At this level,
signals become sparser in time and therefore less recognizable; inevitably,
statistically-motivated pulse-finding techniques are required in order
to separate real pulses from noise or to distinguish them from unrelated
pulses. When signals are not only rare but also extremely small, the
potential for their mis-parametrization (and consequent detection
inefficiency) is high.

The ZEPLIN-III data Reduction and Analysis (ZE3RA) package must therefore
be very accurate at separating small signals from the noise, thus
maximizing sensitivity to WIMPs. It must deal with unexpected rare
artefacts arising in very long measurements; these include internal
effects (e.g. occasional micro-discharges), changes in the local underground
environment (pressure, temperature, structural movements), long-term
electronic drifts and occasional electromagnetic pick-up, and many
others. It must also be very flexible since our knowledge of these
effects improves as more data are accumulated. For these reasons we
opted for deferring the physics interpretation of the detector response
to a later stage in the data analysis. ZE3RA provides robust reduction
of the raw data acquisition output in the form of full pulse parametrization.
Other requirements include a powerful event display and the possibility
to deal with blind analyses, whereby specific events or datasets present
in the data repository should not be displayed or reduced until the
final analysis stage so as not to bias the signal estimation.

In the context of direct WIMP searches, the ultimate benchmark of
such a package is the level of electron/nuclear recoil discrimination.
This is quantified by the relative number of events belonging to the
heavily populated electron-recoil background population (mainly gamma-rays,
with high ionization/scintillation ratios) that leak into the acceptance
region of the signal (mimicking nuclear recoils with low such ratios).
ZEPLIN-III achieved the best discrimination of any two-phase xenon
detector reported so far and, in spite of a relatively high gamma-ray
background, achieved a sensitivity for WIMP signals amongst the best
in the world \citep{Lebedenko:2008aa,Lebedenko:2009mm,Akimov2010180}.
This relied in part on the ability of ZE3RA to parametrize events
correctly. Some of the algorithms contained within it can be applied
to other detector arrays, not only in the context of underground experiments.

\section{Setup and Data Processing\label{sec:Setup}}

ZEPLIN-III is a two-phase (liquid/gas) xenon time projection chamber
containing $\approx12$~kg of liquid xenon above a compact hexagonal
array of $31$ $2$-inch photomultipliers (ETL D766QA) \citep{araujo:2006aa,Akimov:2007aa,Lebedenko:2008aa}.
The photomultipliers (PMTs) are immersed directly in the liquid at
a temperature of $-105{\rm }$$^{o}$C and record both the rapid scintillation
signal (S1) and a delayed second signal (S2) produced by proportional
electroluminescence in the gas phase from charge drifted out of the
liquid \citep{Dolgoshein:1970}. The electric field in the active
xenon volume is defined by a cathode wire grid $36$~mm below the
liquid surface and an anode plate $4$~mm above the surface in the
gas phase. These two electrodes define a drift field in the liquid
of $\approx4$~kV/cm and an electroluminescence field in the gas
of $\approx8$~kV/cm. A second wire grid is located $5$~mm below
the cathode grid just above the PMT array. This grid defines a reverse
field region which suppresses the collection of ionization charge
for events just above the array and helps to isolate the PMT input
optics from the strong external electric field.

\begin{figure}[tbh]
\begin{centering}
\includegraphics[width=1\textwidth]{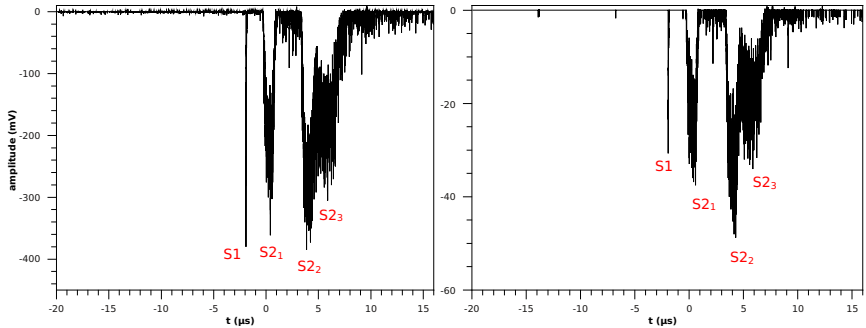}
\par\end{centering}

\caption{\label{fig:event205}Sum of all HS (left) and LS (right) channels
for a low energy event having $3$ energy depositions in the liquid
xenon sensitive volume. The distance between the fast S1 signals (all
overlapping in time) and the corresponding S2 is a measure of the
depth of the interaction. For each S2 pulse the light distribution
over the PMT array allows to reconstruct the position of the interaction
in the (x,y)-plane.}
\end{figure}

The PMT signals are digitized at $2$~ns sampling over a time segment
of $36$~$\mu$s starting at $-20$~$\mu$s from the trigger point.
Each PMT signal is fed into two $8$-bit digitizers (ACQIRIS DC265)
with a $\times10$ gain difference between them provided by fast amplifiers
(Phillips Scientific 770), to obtain both high (HS) and low (LS) sensitivity
readout covering a wide dynamic range. The PMT array is operated from
a common HV supply with attenuators (Phillips Scientific 804) used
to normalize their individual gains. The trigger is generated using
the shaped sum of the HS signals from all the PMTs. For the sake of
illustration, Fig.~\ref{fig:event205} shows the sums of all HS and
LS channels for a low energy multiple scattering event triggered by
the first S2 signal.

\subsection{Software architecture}

\begin{figure*}[tp]
\begin{centering}
\includegraphics[width=1\textwidth]{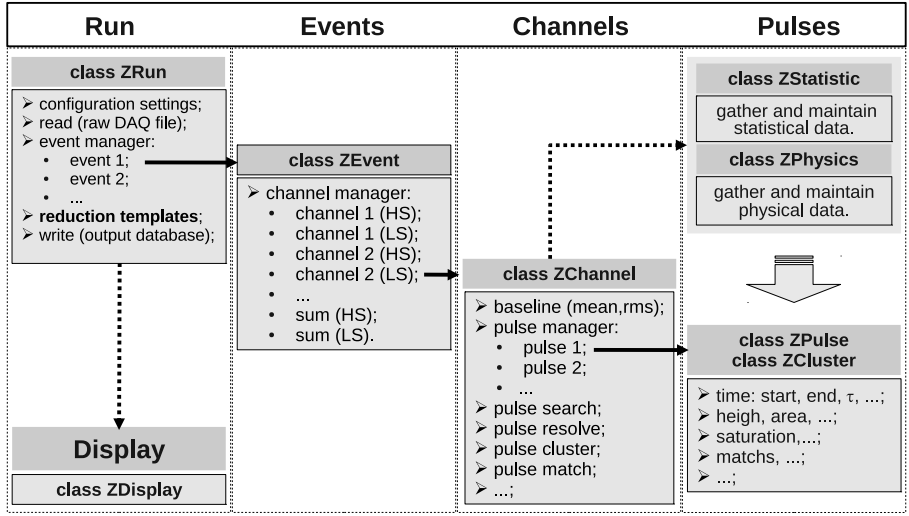}
\par\end{centering}

\caption{\label{fig:schematic}Schematic representation of the ZE3RA software
architecture.}
\end{figure*}

The efficient analysis of the huge amount of data produced by the
acquisition system (DAQ) demands its previous reduction to a set of
relevant physical parameters (pulse timing estimators, height, area,
etc). In the ZEPLIN-III experiment this task is performed by the ZE3RA
software. ZE3RA is implemented in C++ following a comprehensive class-oriented
architecture mimicking the various DAQ functional stages (run, event,
channel, etc). The software design allows easy plug-in of new tools
and to build different reduction templates targeting specific analyses.
These features are of key importance for the optimization of the analysis
which should produce consistent results over a large variety of acquisition
scenarios lasting from a few hours to several months (different field
configurations, calibration sources, etc). The ZE3RA architecture,
illustrated schematically in Fig.~\ref{fig:schematic}, includes
the skeleton classes:
\begin{itemize}
\item The \textit{ZDisplay} class layers the end user interaction with the
classes managing the analysis and holding both the raw and the reduced
data structures.
\item The \textit{ZRun} class inherits from \textit{ZEvent} and manages
all the settings and the different reduction templates.
\item The \textit{ZEvent} class stores and manages the access to individual
events and folded data structures (i.e.\textit{ ZChannel}). It inherits
from \textit{ZRawFileHandler} and \textit{ZNtupleFileHandler} which
layer, respectively, the input from the raw DAQ data files and the
output to the databases containning the reduced quantities. The \textit{ZRawFileHandler}
class inherits from \textit{ZBlindManager} which implements the access
policy to the raw data used in the WIMP search .
\item The \textit{ZChannel} class stores and manages the data from individual
channels and all contained structures (i.g. \textit{ZPulse}).
\item The \textit{ZPulse} and \textit{ZCluster} classes store all information
related with individual pulses and clusters of sequential pulses,
respectively. It inherits from \textit{ZStatistic} and \textit{ZPhysics}
which gather and maintain, respectively, statistical and physical
data.
\end{itemize}
In the following subsections we present a short overview of the most
relevant algorithms implemented and tested in the ZE3RA analysis framework
for the ZEPLIN-III experiment. It should be noted that not all of
those algorithms were used for the production of published results,
either for the first or second science runs (in 2008 and 2010/11,
respectively). In particular, the moving average algorithm (\S\ref{sub:Moving-average-variable})
was preferred to the more refined wavelet analysis (\S\ref{sub:Wavelet-analysis})
as a trade-off between speed and performance. Rather than lessening
the purpose of the software package, this puts the emphasis on its
design and architecture as a general framework easily extensible to
different acquisition scenarios and similar experiments with specific
needs.

\subsection{Baseline characterization.\label{sub:baseline}}

The baseline is parametrized using the waveforms containing the actual
PMT signals. To avoid any bias due to the occurrence of transients
or small spurious signals, the parametrization method relies on a
consistency check of the noise distribution variance during a sufficiently
large time window. For that purpose, the DAQ \textit{pre}-trigger
region is divided into $i=1..M_{0}$ consecutive regions containing
$m$ samples each. For each of these regions, the variances $\left\{ \sigma_{i}^{2},\, i=1..M_{0}\right\} $
of the signal amplitude distribution are calculated. The \textit{F-distribution}
probability function ($Q$) is used to check if the variances are
statistically consistent:
\begin{equation}
Q=\frac{\Gamma(2\nu)}{2\Gamma(\nu)}\int_{0}^{\left(1+F\right)^{-1}}(t-t^{2})^{\nu-1}dt\ ,\label{eq:Q}
\end{equation}
where $\nu=(m-1)/2$ and

\[
\left\{ \begin{array}{ll}
F=\sigma_{i}^{2}/\sigma_{i+1}^{2}, & \sigma_{i}>\sigma_{i+1}\\
F=\sigma_{i+1}^{2}/\sigma_{i}^{2}, & \sigma_{i}\leq\sigma_{i+1}
\end{array}\right.\ .
\]
$Q$ is therefore the significance level at which that hypothesis
($\sigma_{i}^{2}\equiv\sigma_{i+1}^{2},\, i=1..M_{0}$) can be rejected
\citep{Recipes:2002aa}. For each of the $M_{0}$ regions, the means
$\left\{ \mu_{i}^{2},\, i=1..M_{0}\right\} $ of the signal amplitude
distribution are also calculated. The noise \textbf{(}$\sigma_{bas}$)
and mean ($\mu_{bas}$) characterizing each waveform are then defined
as 
\begin{equation}
\sigma_{bas}=\left\langle \sigma_{i}\right\rangle ,\, i=1..M\ ,\label{eq:sigma}
\end{equation}

\begin{equation}
\mu_{bas}=\left\langle \mu_{i}\right\rangle ,\, i=1..M\ ,\label{eq:mean}
\end{equation}
for those $M$ regions satisfying $Q<Q_{crit}$. For the ZEPLIN-III
analysis the values of $Q_{crit}=0.0001$ and $m=25$ ($50$~ns)
were used. The maximum length of the total sampled waveform was $2$~$\mu$s
($M_{0}=40$). For each event the $\sigma_{bas}$, $\mu_{bas}$ and
$M$ values are stored for all channels and can be used, for example,
to identify misbehaving baselines.

\subsection{Raw data filtering.\label{sub:filtering.}}

In order to enhance the signal-to-noise ratio and help with the identification
of relevant pulse structures, a set of general filter algorithms are
available in ZE3RA. Besides the built-in filters, which are briefly
described below, the software framework allows easy plug-in of new
algorithms and their use at any defined configuration. It should be
noted that the DAQ raw data is never modified during the analysis;
instead, an auxiliary buffer containing the filtered data is maintained
for every channel.

\subsubsection{Moving average.\label{sub:Moving-average.}}

The implemented moving average algorithm is a simple low-pass filter
defined as

\begin{equation}
\overline{y}_{i}=\frac{\sum_{k=i-m}^{i+m}y_{k}}{2m+1}\,,\label{eq:mov_average}
\end{equation}
where $\overline{y}$ and $y$ represent, respectively, the filtered
and raw data buffers. Although appealing because of its speed, the
moving average produces a significant loss of information when pulses
are narrow compared with the filter width ($2m+1$). The loss of information
or even distortion of the filtered signal ($\overline{y}$) is an
effect of the weight given to the $k$-th data point in Eq.~\ref{eq:mov_average}
being independent of its distance to $i$. In ZEPLIN-III this presents
a problem since the S1 and S2 signals have very different time constants,
respectively, $\sim30$~ns and $\sim0.6$~$\mu$s (parametrized
by the signal mean arrival time). If $m$ is tuned for S2 then S1
information is lost. Conversely, setting $m$ to preserve S1 does
not help with the detection of S2 pulses. These issues related with
the time scale at which one is filtering or looking for a signal are
addressed by both the methods described in \S\ref{sub:Moving-average-variable}
and \S\ref{sub:Wavelet-analysis}.

\subsubsection{Moving average with variable width.\label{sub:Moving-average-variable}}

A solution to the problem highlighted above, which retains most of
the simplicity and speed of the moving average algorithm, is to adapt
the width of the filter (Eq.~\ref{eq:mov_average}) based on some
characterization $Q_{y}(i,m_{i})$ of the local data being analysed:

\begin{equation}
\overline{y}_{i}=\frac{\sum_{k=i-m_{i}}^{i+m_{i}}y_{k}}{2m_{i}+1}\,,\label{eq:adap_average}
\end{equation}
with

\[
\left\{ \begin{array}{ll}
m_{i+1}=m_{i}+1, & Q_{y}(i,m_{i})<Q_{thr}\bigwedge m_{i}<m_{max}\\
m_{i+1}=m_{i}-1, & Q_{y}(i,m_{i})\geq Q_{thr}\bigwedge m_{i}>m_{min}
\end{array}\right.,
\]
where $m_{min}$, $m_{max}$ and $Q_{thr}$ are user-defined parameters
controlling the limits and the adaptation of the filter width. In
the ZEPLIN-III analysis $Q_{y}(i,m_{i})$ was set to the variance
of $y$ calculated in the $[i-m_{i},i+m_{i}]$ interval. Typical values
of $m_{min}\sim35$ ($70$~ns), $m_{max}\sim100$ ($0.2$~$\mu$s)
and $Q_{thr}^{1/2}=3\sigma_{bas}$ (Eq.~\ref{eq:sigma}) show good
performance in terms of both S1 and S2 pulse detection. Figures \ref{fig:event1}
and \ref{fig:event2} show examples of applying this filter to typical
ZEPLIN-III events and it comparison with the results obtained using
the algorithm described in \S\ref{sub:Wavelet-analysis}.

\subsubsection{Fourier analysis.\label{sub:Fourier-analysis.}}

Besides the random noise intrinsic to any signal processing chain,
in ZEPLIN-III one can also observe the occurrence of coherent noise.
This noise is most often induced by electric equipment working in
close vicinity to the detector or following the saturation of the
amplifiers. In both cases the noise exhibits time periodicity and
is composed of a small set of characteristic frequencies. These properties
make the Fast Fourier Transform (FFT) analysis particularly suitable
for the identification and removal of such occurrences. 

The FFT is an effi{}cient computational tool to calculate the \textit{Fourier}
transform of a function ($y$) sampled at a finite number of $N$
points $\left\{ y_{n},\, n=0..N-1\right\} $ \citep{Recipes:2002aa}.
In the frequency domain the amplitude of the component $f_{n}=n/N\Delta$
is given by

\begin{equation}
H(f_{n})=\Delta\sum_{k=0}^{N-1}y_{k}e^{i2\pi kn/N}=\Delta\sum_{k=0}^{N-1}\left[\cos\left(\frac{2\pi kn}{N}\right)-i\sin\left(\frac{2\pi kn}{N}\right)\right]\,,\label{eq:FFT}
\end{equation}
where $\Delta$ represents the sampling time interval. In ZE3RA the
coefficients $H(f_{n})$ are calculated using the FFTW library \citep{FFTW:2005}.

This FFT tool was used only on a dedicated dataset acquired to study
single electron emission from the liquid into the gas phase \citep{SE-Z3:2011}.
For the acquisition of that dataset the DAQ was triggered externally
with a pulse generator which accidentally induced coherent noise into
the HS channels. After calculating the FFT coefficients for each timeline,
a \textit{$10$-pole Butterworth} filter was used to attenuate $H(f_{n})$
(Eq.~\ref{eq:FFT}) corresponding to the set of noise frequencies
$\left\{ f_{noise}\right\} $ contained in each channel. This procedure
allowed successful recovery of the signal as shown by the comparison
of the single electron emission results obtained using this particular
dataset with the results from another method described in Ref.~\citealp{SE-Z3:2011}
using WIMP search data sets.

\subsubsection{Wavelet analysis.\label{sub:Wavelet-analysis}}

Resulting from being encoded both in terms of amplitude and phase
of sines or cosines (Eq.~\ref{eq:FFT}), the FFT analysis gives no
direct information about the time occurrence of transients \citep{Recipes:2002aa}.
One solution to this problem would be to divide the entire domain
into small regions and analyse them separately. Nevertheless, this
method would imply the loss of information at lower frequencies when
compared with the duration of the region being analysed. Another solution
allowing to preserve both time and frequency information is to use
the Discrete Wavelet Transform (DWT) analysis \citep{Recipes:2002aa}.

The DWT scales ($s$) and shifts ($\tau$) a mother function $\mathcal{F}$
along the time domain while recording its level of correlation with
the signal into a set of coefficients $w(s,\tau)$: 

\begin{equation}
w(s,\tau)=\sum_{i=-\infty}^{+\infty}y_{i}*\mathcal{F}_{(s,\tau),i}\,,\label{eq:wave_forwd}
\end{equation}
where $y$ represents the raw data stream and $\mathcal{F}_{(s,\tau)}$
are the set of basis functions obtained from scaling and translating
$\mathcal{F}$ at fixed $(s,\tau)$ steps. Unlike the sinusoidal functions
which define a unique FFT of the signal, there are many possibilities
for $\mathcal{F}$ producing different DWT coefficients. The choice
of $\mathcal{F}$ is based on the trade-off between localization/smoothing
(time/frequency domain) required for a particular application. The
ZE3RA framework simply wraps the DWT code available from GSL \citep{GSL}
and expands its functionality to facilitate the manipulation of the
coefficients $w(s,\tau)$ (Eq.~\ref{eq:wave_forwd}). The available
$\mathcal{F}$ family functions are: Haar, Daubechies and biorthogonal
b-spline \citep{GSL}.

Currently ZE3RA provides two different noise removal and smoothing
algorithms based on DWT decomposition. The first of these algorithms
implements the soft threshold technique described in Ref.~\citealp{Donoho1995}
for a uniformly distributed \textit{Gaussian} noise: for a given scale
$s$ the $N_{S}$ wavelet coefficients are translated towards $0$
by an amount 

\begin{equation}
\delta_{s}=\frac{\sqrt{2\ln N_{s}}}{0.6745}MAD\,,\label{eq:soft_thresh}
\end{equation}
where $MAD$ represents the median absolute deviation of $\left\{ w_{j}(s,\tau)\, j=1..N_{s}\right\} $.
The second algorithm combines data smoothing with the edge detection
and preservation method described in \citep{Mallat:1992aa}. The edge
detection is based on the multiscale behavior of the local maxima
of the wavelet coefficients moduli (Eq.~\ref{eq:wave_forwd}). The
value of $|w(s,\tau)|$ measures the derivative of the smoothed signal
at the scale $s$ and a signal sharp variation (e.g. S1-like pulses)
produces moduli maxima at different scales \citep{Mallat:1992aa}.
After calculating the moduli of the wavelet coefficients and finding
local maxima, the implemented algorithm maps and stores the inter-scale
evolution of the $|w(s,\tau)|$ values. This information can be used
to study the Lipschitz regularity of the signal and further selection
of the edge/pulse types to retain \citep{Mallat:1992aa}. The signal
coefficients $w(s,\tau)$ which do not belong to a valid edge structure
can either be smoothed using the soft threshold method described above
(Eq.~\ref{eq:soft_thresh}) or simply reset to $0$ at selected higher
order scales (higher frequencies). An additional benefit from using
the second algorithm to smooth the data is the intrinsic availability
of the time position of the edges even in the occurrence of a baseline
drift. This information can later be matched against the algorithm
described in \S\ref{sub:find_pulse} to enhance the pulse finding
efficiency.

\begin{figure}[tbh]
\begin{centering}
\includegraphics[width=1\textwidth]{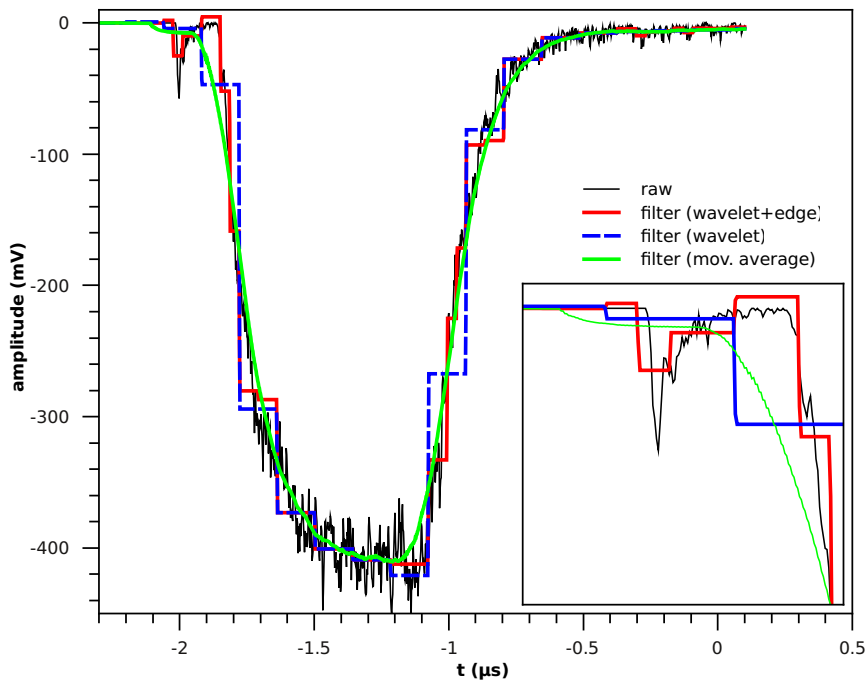}
\par\end{centering}

\caption{\label{fig:event1}Comparison of the smoothing results obtained using
the moving average with variable width (green line) and a DWT decomposition
keeping (red line) or ignoring (blue line) the edge information. This
event corresponds to an interaction just below the liquid xenon surface.
In this instance the small S1 pulse, shown also in the inset, proceeds
a much larger S2 by a very short time.}
\end{figure}

\begin{figure}[tbh]
\begin{centering}
\includegraphics[width=1\textwidth]{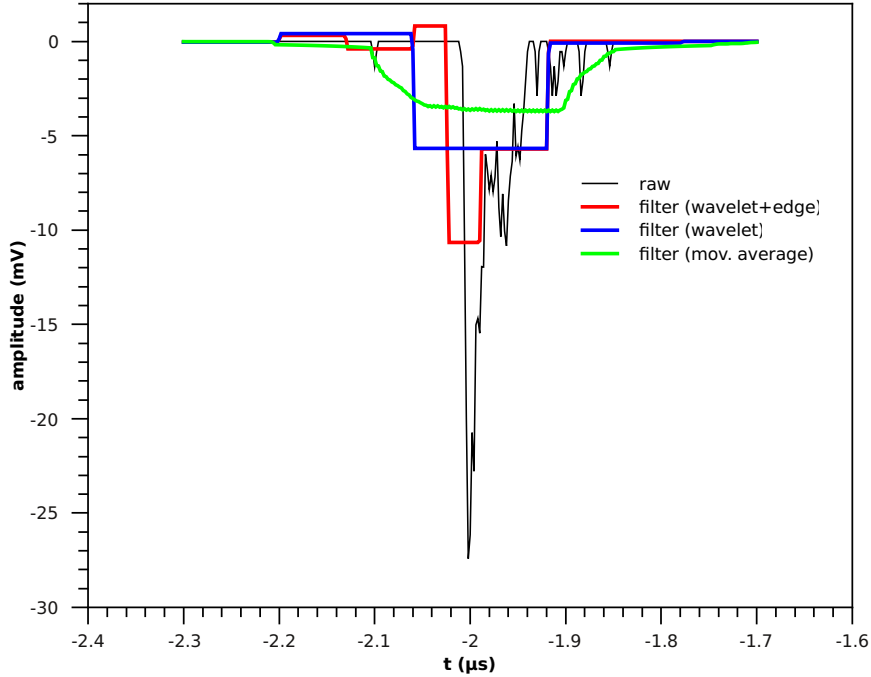}
\par\end{centering}

\caption{\label{fig:event2}Comparison of the smoothing results for a typical
S1 pulse obtained using the moving average with variable width (green
line) and a DWT decomposition keeping (red line) or ignoring (blue
line) the edge information.}
\end{figure}

Figures \ref{fig:event1} and \ref{fig:event2} show the comparison
between the results of the wavelet analysis with edge detection and
of the moving average with adaptive width (\S\ref{sub:Moving-average-variable})
when applied to two typical events. The DWT decomposition was done
into $14$ scales using a bi-orthogonal b-spline mother function ($\mathcal{F}$)
of order $(1,3)$ \citep{GSL}. For the sake of illustration, both
figures show the result of clearing all coefficients above the $8$th
scale ($w(s\geq8,\tau)=0$) together with the effect of keeping those
up to $s=10$ when belonging to an identified S1-like edge. Besides
improving the sensitivity to the general shape and start time of the
fast S1 signals (Fig.~\ref{fig:event2}), smoothing the data using
the edge detection algorithm also improves the discrimination between
S2 and S1 when they are very close (Fig.~\ref{fig:event1}). The
latter class of events corresponds to energy depositions near the
surface of the liquid.

\subsection{Channel delays.\label{sub:Channel-delays}}

Due to differences in the signal processing chain (PMTs, cabling,
amplifiers, etc), all ZEPLIN-III channels exhibit relative delays
ranging typically from $\pm10$~ns. The correct alignment in time
of all channels is crucial for the performance of both the pulse finding
and matching algorithms (\S\ref{sub:find_pulse} and \S\ref{sub:match_pulse},
respectively). This is specially relevant for signals corresponding
to low energy deposits, which constitute the region of interest in
ZEPLIN-III.

The individual channels are realigned in ZE3RA by defining the beginning
of the raw data buffer $y^{(i)}$ to point at the $k$-th element
of the buffer $Y^{(i)}$ containing the original data read from the
DAQ, 

\begin{equation}
k_{Y}^{(i)}=\delta^{(i)}-\min\left\{ \delta^{(j)},\, j=1..J\right\} \,,\label{eq:rebegin}
\end{equation}
where $\delta^{(i)}$ is the individual delay for channel $i$ and
$J$ represents the total number of channels. The size $S_{y}$ for
all $y$ buffers is calculated using

\begin{equation}
S_{y}=S_{Y}+\min\left\{ \delta^{(j)},\,1..J\right\} -\max\left\{ \delta^{(j)},\, j=1..J\right\} \,,\label{eq:resize}
\end{equation}
where $S_{Y}$ is the size of the original $Y$ buffers. The very
simple operations defined in Eq.~\ref{eq:rebegin} and Eq.~\ref{eq:resize}
avoid any extra time- or memory-consuming manipulations aside from
reading the DAQ files into buffers $\left\{ Y^{(j)},\, j=1..J\right\} $.

\subsection{Pulse finding.\label{sub:find_pulse}}

\begin{figure}[tbh]
\begin{centering}
\includegraphics[width=1\textwidth]{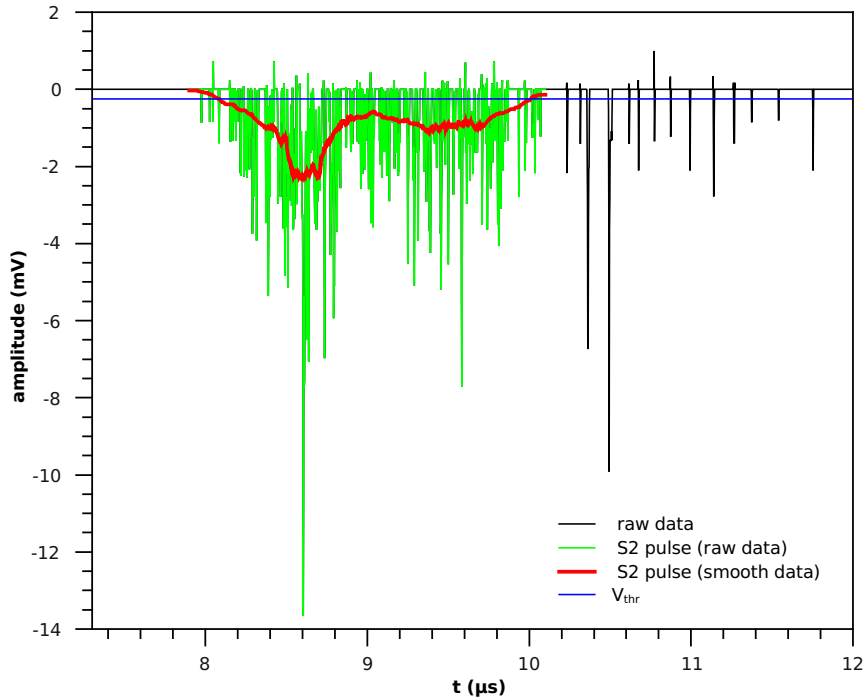}
\par\end{centering}

\caption{\label{fig:pulse835}Example of a ZEPLIN-III event with a S2 pulse
(green line) being selected from an excursion of $\overline{y}$ (red
line) above $V_{thr}$ (blue line). The fastest pulses on the right
tail of the S2 are chosen from $y$ (black line).}
\end{figure}

The pulse finding procedure implemented in ZE3RA consists of searching
for excursions of the signal amplitude above a defined threshold $V_{thr}$.
However, in the ZEPLIN-III detector both S1 and S2 signals can contain
several of these excursions depending on the distance from a particular
PMT to the interaction point and on the energy deposited. In particular,
an S2 signal can consist of a few tens of photoelectrons spread over
a period of time of $\sim1$~$\mu$s. Taking advantage of the underlying
structure revealed from filtering (\S\ref{sub:filtering.}), this
problem was partially solved in ZE3RA by first searching for pulses
on the smooth data buffer ($\overline{y}$). One must keep in mind
though that the effective enhancement of smoothing depends on the
applied filter and how sparsely the individual data excursions occur
(\S\ref{sub:filtering.}). Regardless of any loss of information
in $\overline{y}$, ZE3RA keeps the sensitivity to the smallest structures
by also searching for pulses in the original data buffer ($y$). The
final set of pulses available for all subsequent analysis is the union
of pulses collected from both $\overline{y}$ and $y$. For the ZEPLIN-III
analysis $V_{thr}$ was chosen to be

\begin{equation}
V_{thr}=\mu_{bas}+3\sigma_{bas}\,,\label{eq:threshold}
\end{equation}
where $\sigma_{bas}$ is the noise (Eq.~\ref{eq:sigma}) and $\mu_{bas}$
the mean (Eq.~\ref{eq:mean}) values characterizing each raw waveform
(\S\ref{sub:baseline}). This software threshold is nominally equivalent
to an energy threshold of only $1.67$~keV for electron recoils detected
through S1. As an example, Figure \ref{fig:pulse835} shows a larger
pulse which is chosen from $\overline{y}$ with the remaining fastest
pulses being picked from the unmodified $y$ buffer.

\subsection{Pulse clustering. \label{sub:cluster-pulse}}

Deciding when to stop accruing small excursions above threshold into
pulse clusters is not straightforward and has important consequences
for the detection efficiency of small signals. For S1 pulses, for
example, a fixed-length integration, typically implemented as a coincidence
window between channels, can lead to unnecessary inclusion of noise
and unequal integration efficiency for different particle species
(due to the different scintillation decay time constants). Alternatively,
one may cluster subsequent candidates into the pulse based on time
separation, although over-clustering can lead to run-away effects
in this instance. The fraction of the pulse area integrated in each
approach can be calculated analytically if the scintillation responses
are known for each species, but this calculation fails for very low
photoelectron numbers, when the start time of the pulse is not defined
by the rise time of the scintillation signal but rather by the delayed
arrival of the first photoelectron. A detailed comparison between
the constant integration and the pulse gap methods was carried out
with a toy \textit{Monte Carlo} accounting for the DAQ sampling rate,
the width of the single photoelectron response and the scintillation
decay times and respective intensities for electron and nuclear recoils
in liquid xenon \citep{Kwong:2010aa} (however, noise or afterpulsing
distributions were not included). This was used to calculate the fraction
($\eta$) of S1 signals which is lost from the integrated pulse area.
The integration of S2 pulses is not so critical as $\eta$ is expected
to be always very small in this case.

\begin{figure}[tbh]
\begin{centering}
\includegraphics[width=1\textwidth]{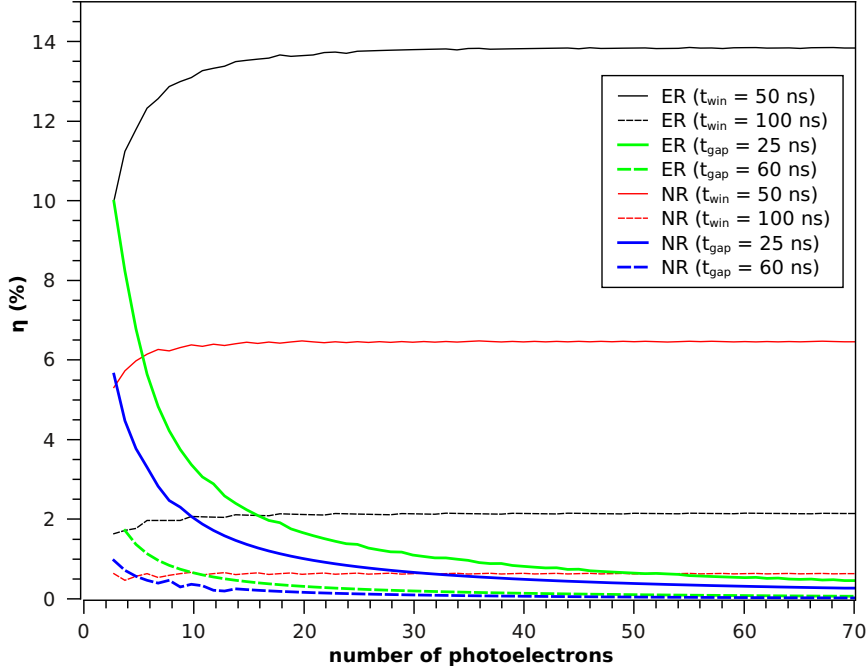}
\par\end{centering}

\caption{\label{fig:cluster_eff}Comparison for both electron (ER) and nuclear
(NR) recoils in the liquid xenon of the fraction of lost area of S1
signal ($\eta$) when clustering pulses using a constant time window
($t_{win}=50,\,100$~ns) or the time gap between pulses ($t_{gap}=25,\,60$~ns).}
\end{figure}

The simplest pulse clustering algorithm implemented in ZE3RA consists
of merging all pulses found within a time window of constant width
($t_{win}$). The value of $t_{win}$ is chosen according to the characteristics
of the signal (i.e, S1 or S2). The respective \textit{Monte Carlo}
results are shown in Fig.~\ref{fig:cluster_eff} as a function of
the sampled number of photoelectrons for window sizes of $50$~ns
and $100$~ns. As expected, the missing area fraction $\eta$ is
smaller for nuclear recoils (due to the increase of the faster xenon
scintillation component); significantly, $\eta$ is not constant for
small signals, but rather it decreases due to the delayed detection
of the first photoelectron, which is a purely statistical effect.

The alternative clustering algorithm implemented uses the time distance
between pulses to decide if they correspond to the same interaction
in the liquid xenon. The algorithm iterates through pulses and recursively
merges consecutive occurrences if the time elapsed between the end
of the first and the begin of the next is smaller then a certain value
$t_{gap}$. It should be noted that the saturation tails from the
amplifiers and the existence of afterpulsing signals originated in
the PMTs \citep{Coates:1973aa}  can bias the results towards excessive
clustering. To mitigate these, one additional constraint is imposed
in ZE3RA: that the clustering should not extend out to more then a
user defined factor of the maximum mean arrival time of photoelectrons
for the pulses being clustered. The\textit{ Monte Carlo} results are
also shown in Fig.~\ref{fig:cluster_eff} for $t_{gap}=25,\,60$
~ns. For pairs of values ($t_{gap},t_{win}$) returning similar values
at the lowest number of sampled photoelectrons, the performance of
the gap method improves quickly with the magnitude of the signal and
almost independently of the type of particle.

Using real detector data, we verified that the time gap algorithm
was more robust than the constant time window option when dealing
with different acquisition scenarios (different fields, detector tilt
or changes in the gas gap, etc) and with the interaction of different
particles in the liquid xenon (from gamma and neutron calibrations).
We also found that the constant integration method, with $t_{win}$
tuned for specific conditions and single scatters in the liquid, often
clipped multiple overlapping pulses, therefore biasing their parametrization
and correct identification. These reasons, and the potential for better
performance on rare pulse topologies which may arise in very long
exposures, led us to adopt the time difference method, with a simple
heuristic adjustment of $t_{gap}$ to $20$~ns and $100$~ns for
S1- and S2-like pulses, respectively.

\subsection{Multiple pulse resolution.\label{sub:pulse-resolution.}}

In the ZEPLIN-III experiment only events with one S1 and one S2 are
considered for most analysis (e.g. WIMPs search). Events corresponding
to multiple scatters can be promptly identified using the S2 channel
if the individual energy deposits occur at different depths in the
LXe active volume. The time separation between the S1 and S2 signals
is equal to time taken by the ionization electrons to travel from
the interaction site to the liquid surface along the drift field.
This mechanism is independent of the position of the interactions
in the $(x,y)$-plane which, in any case, can only be obtained using
position reconstruction after the correct identification and parametrization
of the pulses.

\begin{figure}[tbh]
\begin{centering}
\includegraphics[width=1\textwidth]{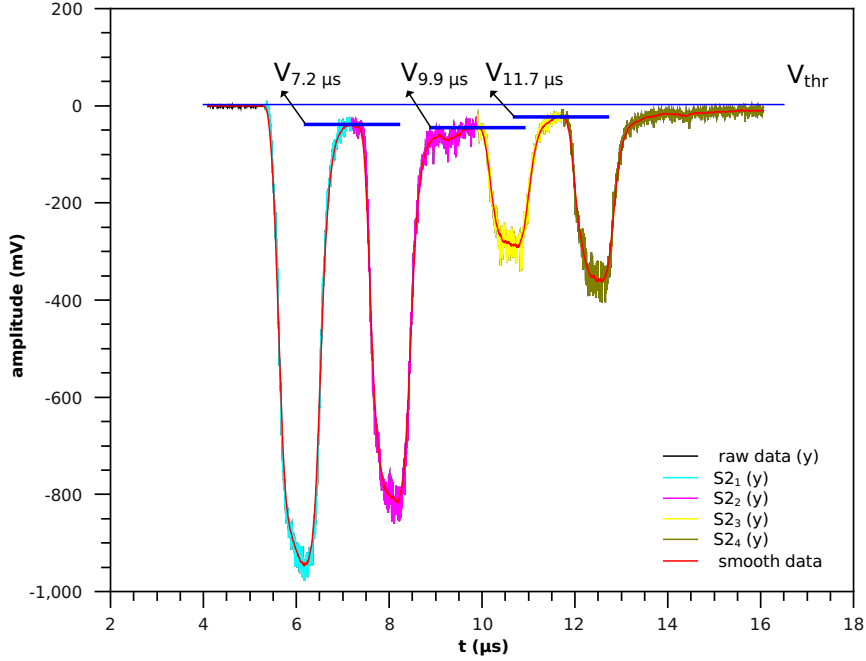}
\par\end{centering}

\caption{\label{fig:event145}ZEPLIN-III event with $4$ overlapping S2 pulses
(light blue, magenta, yellow and olive). The figure also shows the
set of thresholds $\left\{ V_{11.7\,\mu s},V_{7.2\,\mu s},V_{9.9\,\mu s}\right\} $
gathered from scanning $\overline{y}$ (red line) and used to resolve
the original pulse found from an excursion of the data past $V_{thr}$.}
\end{figure}

The algorithm implemented in ZE3RA to identify and resolve multiple
scatter events with overlapping S2 signals consists simply of reusing
the pulse finding and clustering algorithms described in \S\ref{sub:find_pulse}
and \S\ref{sub:cluster-pulse} for higher thresholds. For each S2
a list of thresholds is obtained by scanning the smooth buffer ($\overline{y}$)
and accumulating the values $V_{l}=\overline{y}_{l}+V_{thr}$ (Eq.~\ref{eq:threshold})
obeying:

\begin{equation}
\overline{y}_{k}-\overline{y_{l}}>V_{thr}\,\wedge\,\overline{y}_{m}-\overline{y}_{l}>V_{thr},\, k<l<m\,,\label{eq:thr_resolve}
\end{equation}
where $k$ and $m$ are constrained by the candidate pulse start and
end times $(t_{start},t_{end})$. The pulse finding and clustering
algorithms are then applied to the data sub-domain defined by $(t_{start},t_{end})$
using the lowest value in $\left\{ V_{l}\right\} $. The algorithm
recursively applies this method to every resolved S2 pulse using the
remaining threshold values in the $V_{l}$ list. To illustrate the
procedure, Figure \ref{fig:event145} shows the set of test thresholds
$\left\{ V_{11.7\,\mu s},V_{7.2\,\mu s},V_{9.9\,\mu s}\right\} $
for a ZEPLIN-III event with $4$ overlapping S2 pulses.

\subsection{Pulse matching.\label{sub:match_pulse}}

One frequent requirement of experiments using an array of photodetectors
is to order in time and match signals from all DAQ channels. The simplest
approach would be to have the ordering and matching function $O$
depending both on the start and end times $(t_{start},t_{end})$ of
any two pulses $\left\{ A,B\right\} $ occurring in different channels
$\{a,b\}$,

\begin{equation}
O(A,B)=\begin{cases}
\begin{array}{c}
A\,\text{precedes}\, B\\
A\,\text{follows}\, B\\
A\,\text{matches}\, B
\end{array} & \begin{array}{c}
,\, t_{end}^{(A)}<t_{start}^{(B)}\\
,\, t_{start}^{(A)}>t_{end}^{(B)}\\
,\,\text{otherwise}
\end{array}\end{cases}\,.\label{eq:match}
\end{equation}
An obvious scenario where the above method fails is when the pulse
limits are extended out due to the occurrence of some sort of noise,
amplifier saturation tail, etc. The possibility of having multiple
interactions with partially overlapping S2s can also drive Eq.~\ref{eq:match}
to faulty results (\S\ref{sub:pulse-resolution.}). In order to solve
these problems, an algorithm was implemented which takes into account
a relative quantification $W$ of the matching between pulses,

\begin{equation}
W(A,B)=\sum_{i\in\left\{ A\cap B\right\} }y_{i}^{(a)}*y_{i}^{(b)}\,,\label{eq:weight}
\end{equation}
where $y^{(a)}$ and $y^{(b)}$represent, respectively, the data streams
containing pulses $A$ and $B$. The algorithm starts by using Eq.~\ref{eq:match}
to order and group overlapping pulses from the two channels $\left\{ a,b\right\} $.
For each one of these groups, eq.~\ref{eq:weight} is used to generate
an overall matching quantity $W$ defined as,

\begin{equation}
W=\prod_{k,m}W(A_{k},B_{m})\delta_{km}\,.\label{eq:total_weight}
\end{equation}
The values of $\delta_{km}$ are initially set to $1$ for all pairs
of $(k,m)$ pulses. A fast \textit{best-survival} algorithm was then
implemented which maximizes the value of $W$ while setting $\delta_{km}$
to $0$ (Eq.~\ref{eq:total_weight}) for all but the best degenerated
matches $W(A_{k},B_{m})$ (Eq.~\ref{eq:weight}). The application
of this method in ZE3RA introduces a flavour of most probable match,
even if the true pulse shape is not considered, which increased significantly
the robustness of the pulse finding analysis.

The relation between pulses from different channels generated by the
matching algorithm was used in several of the reduction stages in
order to:
\begin{itemize}
\item ignore LS pulses lacking a match in the respective HS channel (\S\ref{sub:find_pulse});
\item cluster groups of scattered pulses in individual channels matching
only one pulse in the correspondent HS or LS sum channels (\S\ref{sub:pulse-resolution.});
\item disentangle the correspondence between pulses in individual channels
with overlapping pulses in the sum channels for multiple interactions
at different (x,y)-positions (\S\ref{sub:match_pulse});
\item avoid misinterpretation of amplifier saturation tails at individual
channels as multiple pulses in the sum channels;
\item map the correspondence between pulses or groups of pulses (clusters)
in all channels for the output databases containing the reduced quantities
(\S\ref{sub:parameterization.});
\end{itemize}

\subsection{Pulse parametrization.\label{sub:parameterization.}}

Once the pulse start and end times are set resulting from the operations
described in sections \ref{sub:baseline} to \ref{sub:match_pulse},
a number of parameters are extracted from the raw data buffers. The
actual list of parameters can depend on the pulse context information
(e.g. whether the pulse has a saturation or ringing tail, etc) and
is extensible to support any subsequent analysis. For the ZEPLIN-III
analysis the reduced databases include:
\begin{itemize}
\item pulse start time, width (between threshold crossings, at 10\% and
at 50\% height), amplitude, area, signal mean arrival time, pulse
symmetry, etc;
\item pulse matching mapping over different acquisition channels, etc;
\end{itemize}
During any hypothetical step of the analysis it may be convenient
to access some characteristic of an individual pulse prior to its
formal and complete parametrization: consider, for example, that one
requires the area of a pulse before its clustering (\S\ref{sub:cluster-pulse}).
For that purpose, any parameter operator $\alpha$ that applied to
pulses $A$ and $B$ obeys

\begin{equation}
\alpha(A)+\alpha(B)=\alpha(A\cup B)\label{eq:inline_operator}
\end{equation}
is updated every time the pulse start or end times changes. This feature
of the framework increases significantly the performance of the analysis
since it avoids redundant loops over the data buffers for $\alpha$.

\section{Interface.\label{sec:Interface}}

The ZE3RA human interface layers the end-user interaction with the
base analysis framework and can be used either in graphical or batch
operation mode.

The graphical interface mode was designed to allow easy navigation
through events and channels while providing an inline tuning of the
relevant analysis parameters and visualization of their output in
terms of pulse identification and parametrization. To help understanding
of the detector output, the interface also incorporates a \textit{2D}
plot of the relative response of the array, the possibility to plot
together any set of high/low sensitivity channels in any temporal
scale and mouse context information on pulse properties. The interface
was coded using the cross-platform GUI toolkit FLTK (Fast Light Toolkit).
However, it is worth noting that the core analysis classes are self-contained
(class \textit{ZRun}) and independent of the graphical interface.
This provides the ability to promptly reuse the developed analysis
framework together with any available graphical tool or within the
DAQ environment for online monitoring of the detector. Significantly,
it makes the porting of this framework to other detector systems quite
straightforward.

The batch mode consists of a simple command line input designed mainly
to perform the mass reduction of any set of data files with minimum
user interaction (for example, the first science run data consisted
of 13,234 binary files containing over 23 million triggered events).
The analysis input parameters are fed as a configuration file which
can easily be created from the graphical interface.

An important feature affecting both interface modes is the built-in
blind manager. This manager allows a \textit{super-user} to set, based
on a simple set of rules, which events or datasets can be visualized
and reduced at a specific step of the WIMP search analysis. Such signal-blind
analyses are an essential tool of rare rare event searches.

\section{Conclusions}

A robust and versatile software package was developed for the analysis
and reduction of the raw data from the ZEPLIN-III experiment. The
framework allows the easy plug-in of new tools and building of different
reduction templates targeting specific analyses. The very high electron/nuclear
recoil discrimination achieved in the WIMP search carried out in the
first science run (>99.98\%) benchmarks these algorithms. These techniques
should find application in data reduction from detectors with a large
number of channels, beyond the field of rare event searches.

\section{acknowledgments}

The UK groups acknowledge the support of the Science \textbackslash{}\&
Technology Facilities Council (STFC) for the ZEPLIN--III project and
for maintenance and operation of the underground Palmer laboratory
which is hosted by Cleveland Potash Ltd (CPL) at Boulby Mine, near
Whitby on the North-East coast of England. The project would not be
possible without the co-operation of the management and staff of CPL.
We also acknowledge support from a Joint International Project award,
held at ITEP and Imperial College, from the Russian Foundation of
Basic Research (08-02-91851 KO a) and the Royal Society. LIP--Coimbra
acknowledges financial support from Fundação para a Ciência e Tecnologia (FCT)
through the project-grants CERN/FP/109320/2009 and CERN/FP/116374/2010,
as well as the postdoctoral grants SFRH/BPD/27054/2006, SFRH/BPD/47320/2008
and SFRH/BPD/63096/2009. This work was supported in part by SC Rosatom,
contract \$\textbackslash{}\#\$H.4e.45.90.11.1059 from 10.03.2011.
The University of Edinburgh is a charitable body, registered in Scotland,
with the registration number SC005336. 

\bibliographystyle{elsarticle-num}
\addcontentsline{toc}{section}{\refname}\bibliography{/media/FNEVES/library/jabref}

\begin{thebibliography}{10}
\expandafter\ifx\csname url\endcsname\relax
  \def\url#1{\texttt{#1}}\fi
\expandafter\ifx\csname urlprefix\endcsname\relax\def\urlprefix{URL }\fi
\expandafter\ifx\csname href\endcsname\relax
  \def\href#1#2{#2} \def\path#1{#1}\fi

\bibitem{sumner01}
T.~J. {Sumner}, {Proc. 3rd Int. Workshop on the Identification of Dark Matter},
  ed. N. J. C. Spooner and V. Kudryavtsev, Singapore: World Scientific (2001)
  452.

\bibitem{araujo:2006aa}
H.~M. Ara{\'u}jo, et~al., {The ZEPLIN-III dark matter detector: performance
  study using an end-to-end simulation tool}, Astroparticle Physics 26 (2006)
  140.

\bibitem{Akimov:2007aa}
D.~Akimov, et~al., {The ZEPLIN-III dark matter detector: Instrument design,
  manufacture and commissioning}, Astroparticle Physics 27~(1) (2007) 46 -- 60.

\bibitem{Lebedenko:2008aa}
V.~N. Lebedenko, et~al., {Results from the first science run of the ZEPLIN-III
  dark matter search experiment}, Phys. Rev. D (Particles and Fields) 80~(5)
  (2009) 052010.

\bibitem{Lebedenko:2009mm}
V.~N. Lebedenko, et~al., {Limits on the Spin-Dependent WIMP-Nucleon Cross
  Sections from the First Science Run of the ZEPLIN-III Experiment}, Phys. Rev.
  Lett. 103~(15) (2009) 151302.

\bibitem{Akimov2010180}
D.~Akimov, et~al., {Limits on inelastic dark matter from ZEPLIN-III}, Physics
  Letters B 692~(3) (2010) 180 -- 183.

\bibitem{Dolgoshein:1970}
B.~A. {Dolgoshein}, V.~N. {Lebedenko}, B.~U. {Rodionov}, new method of
  registration of ionizing-particle tracks in condensed matter, JETP Lett. 11
  (1970) 351.

\bibitem{Recipes:2002aa}
W.~H. Press, B.~P. Flannery, S.~A. Teukolsky, W.~T. Vetterling, {Numerical
  recipes in C : The art of scientific computing}, {Cambridge University Press,
  2002}.

\bibitem{FFTW:2005}
M.~Frigo, S.~G. Johnson, The design and implementation of fftw3, in:
  Proceedings of the IEEE, no. 93 (2), 2005, pp. 216--231.

\bibitem{SE-Z3:2011}
{ZEPLIN-III}, Single electron emission in the zeplin-iii two-phase xenon
  detector, In Preparation.

\bibitem{GSL}
M.~Galassi, et~al., GNU Scientific Library Reference Manual (3rd Ed.), ISBN
  0954612078.

\bibitem{Donoho1995}
D.~Donoho, De-noising by soft-thresholding, Information Theory, IEEE
  Transactions on 41~(3) (1995) 613 --627.

\bibitem{Mallat:1992aa}
S.~Mallat, S.~Zhong, Characterization of signals from multiscale edges, IEEE
  Transactions on Pattern Analysis and Machine Intelligence 14 (1992) 710--732.

\bibitem{Kwong:2010aa}
J.~Kwong, P.~Brusov, T.~Shutt, C.~Dahl, A.~Bolozdynya, A.~Bradley,
  Scintillation pulse shape discrimination in a two-phase xenon time projection
  chamber, Nucl. Inst. and Meth. A 612~(2) (2010) 328 -- 333.

\bibitem{Coates:1973aa}
P.~B. Coates, The origins of afterpulses in photomultipliers, Journal of
  Physics D: Applied Physics 6~(10) (1973) 1159.

\end{thebibliography}

\end{document}